\documentclass[a4paper,10pt,twoside]{cpc-hepnp}
\usepackage{CJK,upgreek,fancyhdr}
\usepackage{multicol}
\usepackage{graphicx}
\usepackage{booktabs}
\usepackage{amssymb,bm,mathrsfs,bbm,amscd}
\usepackage[tbtags]{amsmath}
\usepackage{lastpage}
\usepackage{longtable}
\usepackage{changes}
\usepackage{lineno,hyperref}
\usepackage{threeparttable} 
\usepackage{multirow}
\usepackage{subfigure} 
\usepackage{ulem}
\usepackage{color}
\usepackage{xcolor}
\usepackage{amsmath}
\usepackage{diagbox}
\usepackage{verbatim}
\usepackage{flushend}
\usepackage{cancel}
\begin{document}
\begin{CJK*}{GB}{gbsn}

\fancyhead[c]{\small Chinese Physics C~~~Vol. xx, No. x (2021) xxxxxx}
\fancyfoot[C]{\small 010201-\thepage}

\title{Two-proton radioactivity within Coulomb and proximity potential model
\thanks{We would like to thank Jun-Gang Deng, Xiao Pan, Hai-Feng Gui for valuable discussions and suggestions of this work. This work was supported by the National Natural Science Foundation of China (Grants No. 12175100 and No. 11975132), the Construct Program of the Key Discipline in Hunan Province, the Research Foundation of Education Bureau of Hunan Province, China (Grant No. 18A237), the Natural Science Foundation of Hunan Province, China (Grant No. 2015JJ3103 and No. 2018JJ2321), the Innovation Group of Nuclear and Particle Physics in USC, the Shandong Province Natural Science Foundation, China (Grant No. ZR2019YQ01), Hunan Provincial Innovation Foundation For Postgraduate (Grant No. CX20210942), the Opening Project of Cooperative Innovation Center for Nuclear Fuel Cycle Technology and Equipment, University of South China (Grant No. 2019KFZ10).}}

\author{%
\quad De-Xing Zhu    (×£µÂÐÇ)$^{1}$
\quad Hong-Ming Liu  (ÁõºêÃú)$^{1}$
\quad Yang-Yang Xu   (ÐìÑîÑó)$^{1}$
\quad You-Tian Zou   (×ÞÓÐÌð)$^{1}$\\
\quad Xi-Jun Wu      (Îâϲ¾ü)$^{4}$\email{wuxijun1980@yahoo.cn}
\quad Peng-Cheng Chu (³õÅô³Ì)$^{5}$\email{kyois@126.com}
\quad Xiao-Hua Li (ÀîС»ª)$^{1,2,3}$\email{lixiaohuaphysics@126.com}\\
}

\maketitle

\address{%
$^1$School of Nuclear Science and Technology, University of South China, Hengyang 421001, People's Republic of China\\
$^2$National Exemplary Base for International Sci $\&$ Tech. Collaboration of Nuclear Energy and Nuclear Safety, University of South China, Hengyang 421001, People's Republic of China\\
$^3$Cooperative Innovation Center for Nuclear Fuel Cycle Technology $\&$ Equipment, University of South China, Hengyang 421001, People's Republic of China\\
$^4$School of Math and Physics, University of South China, Hengyang 421001, People's Republic of China\\         
$^5$Institute of Theoretical Physics, School of Science, Qingdao Technological	University, Qingdao 266000, People's Republic of China\\
}

\begin{abstract}
In the present work, considering the preformation probability of the emitted two protons in the parent nucleus, we extend the Coulomb and proximity potential model (CPPM) to systematically study two-proton ($2p$) radioactivity half-lives of the nuclei close to proton drip line, while the proximity potential is chosen as Prox.81 proposed by Blocki \textit{et al}. in 1981. Furthermore, we apply this model to predict the half-lives of possible $2p$ radioactive candidates whose $2p$ radioactivity is energetically allowed or observed but not yet quantified in the evaluated nuclear properties table NUBASE2016. The predicted results are in good agreement with those from other theoretical models and empirical formulas, namely the effective liquid drop model (ELDM), generalized liquid drop model (GLDM), Gamow-like model, Sreeja formula and Liu formula.
\end{abstract}



\footnotetext[0]{\hspace*{-3mm}\raisebox{0.3ex}{$\scriptstyle\copyright$}2021
Chinese Physical Society and the Institute of High Energy Physics
of the Chinese Academy of Sciences and the Institute
of Modern Physics of the Chinese Academy of Sciences and IOP Publishing Ltd}%

\begin{multicols}{2}
	
\section{Introduction}
\label{section 1}
Two-proton ($2p$) radioactivity was firstly predicted by Zel'dovich in 1960s, followed by the description of this process was given by Goldansky\cite{Sov.Phys.JETP.11.812,Nucl.Phys.19.482,Nucl.Phys.27.684,Nucl.Phys.23.366}. Subsequently, a great deal of efforts on experiments and/or theories are devoted to explore the probable $2p$ radioactivity phenomena, which opens a new window to study the decay modes and ground-state masses of exotic nuclei near or beyond proton drip line\cite{Phys.Rev.C.90.054326,Eur.Phys.J.A.55.214,Phys.Rev.C.96.034619,Chin.Phys.C.45.044110,Phys.Rev.C.79.054330,Chin.Phys.C.45.104102,Phys.Rev.C.68.054005,Phys.Rev.Lett.111.139903,Phys.Rev.C.100.054332,Eur.Phys.J.A.55.33,Eur.Phys.J.A.54.65,Phys.Rev.C.64.054002,Commun.Theor.Phys.73.075302,Int.J.Mod.Phys.E.30.2150074,Commun.Theor.Phys.73.075301}. Moreover, the study of $2p$ radioactivity can extract abundant nuclear structure information, such as the sequences of particle energies, the wave function of emitted two protons, deformation effect and so on\cite{Rep.Prog.Phys.71.046301,Phys.Rev.C.69.054311,Prog.Part.Nucl.Phys.59.418}. However, by reason of the limitations of experimental techniques, it is extremely difficult to observe the $2p$ radioactivity phenomenon from nuclear ground state in the early experiments. With the development of experimental facilities and detection technologies, the not true $2p$ radioactivity ($Q_{2p}\, \textgreater\,0$ and $Q_p\,\textgreater\,0$, where $Q_p$ and $Q_{2p}$ are the released energy of proton radioactivity and two-proton radioactivity, respectively) were observed from a very short-lived nuclear ground state, such as $^6$Be\cite{Phys.Rev.150.836}, $^{12}$O\cite{Phys.Rev.Lett.74.860} and $^{16}$Ne\cite{Phys.Rev.C.17.1929}. In 2002, the true $2p$ radioactivity ($Q_{2p}\,\textgreater\,0$ and $Q_p\,\textless\,0$) \cite{Rev.Mod.Phys.84.567} was discovered from ground state of $^{45}$Fe at GSI\cite{Eur.Phys.J.A.14.279} and GANIL\cite{Phys.Rev.Lett.89.102501}, respectively. Whereafter, a series of true $2p$ radioactivity phenomena were also detected, such as $^{54}$Zn\cite{Phys.Rev.Lett.94.232501,Phys.Rev.Lett.107.102502}, $^{19}$Mg\cite{Phys.Rev.Lett.99.182501} and so on\cite{Phys.Rev.Lett.117.162501,Phys.Rev.Lett.84.1116,Phys.Rev.C.83.061303}.

From the theoretical point of view, several approaches have been proposed to analyze the $2p$ radioactivity during the recent decades\cite{J.Phys.G:Nucl.Part.Phys.31.S1509,Phys.Rev.Lett.111.139903,Eur.Phys.J.A.54.65,Eur.Phys.J.A.48.179,Phys.Rev.C.54.1240,Phys.Rev.C.56.1866,Phys.Rev.C.58.2813,Phys.Rev.C.59.726,Phys.Rev.C.53.214}, which can be roughly divided into two categories. The one considers the emitted two protons from the parent nucleus being correlated strongly and formed a $^2$He-like cluster, including the effective liquid drop model (ELDM)\cite{Phys.Lett.B.774.14}, Gamow-like model\cite{Chin.Phys.C.45.044110}, generalized liquid drop model (GLDM)\cite{Phys.Rev.C.101.014301}, etc. The other refers to the emitted two protons process being an isotropic emission with no angular correlation, which treated the parent nucleus is composed by two protons and a remnant core that usually called three body radioactivity\cite{Phys.Rev.Lett.111.139903,Phys.Rev.C.68.054005,Nucl.Phys.56.86,Phys.Rev.Lett.110.222501,Phys.At.Nucl.56.460,Phys.Rev.Lett.85.22,Phys.Rev.C.63.034607,Nucl.Phys.A.765.370,Phys.Rev.C.76.014008,Phys.Rev.C.78.034004,Phys.Rev.C.77.064305,Phys.Rev.C.82.034001,Acta.Phys.Pol.B.42.545}. Furthermore, empirical formulas were also proposed to investigate $2p$ radioactivity, such as the four-parameter and two-parameter empirical formulas which were proposed by Sreeja \textit{et al}.\cite{Eur.Phys.J.A.55.33} and Liu \textit{et al}.\cite{Chin.Phys.C.45.024108}, respectively. Within these approaches and/or empirical formulas, the experimental $2p$ radioactivity half-lives are reproduced with different accuracies.

The proximity potential was firstly put forward by Blocki \textit{et al}. based on the proximity force theorem\cite{Ann.Phys.105.427} and widely applied to nuclear physics\cite{Nucl.Phys.A.850.34,Nucl.Phys.A.935.28,Can.J.Phys.95.31,Phys.Rev.C.62.044610,Nucl.Phys.A.438.450,Nucl.Phys.A.922.191,J.Phys.G:Nucl.Part.Phys.41.105108,Nucl.Phys.A.882.49,Nucl.Phys.A.889.29,Nucl.Phys.A.929.20}, such as heavy-ion fusion reaction\cite{Pramana.76.6}, heavy-ion elastic scattering\cite{Phys.Let.65B.1}, fusion barriers\cite{Phys.Rev.C.81.044615}, etc. For its simple and accurate formalism with the advantage of adjustable parameters, using the proximity potential to replace the nuclear potential, Santhosh \textit{et al}. have been proposed the Coulomb and proximity potential model (CPPM)\cite{Pramana.58.611} to deal with cluster radioactivity in 2002. Hence, CPPM\cite{Pramana.58.611} was extended to study $\alpha$ decay\cite{Eur.Phys.J.A.51.122,Phys.Rev.C.93.024612,Phys.Rev.C.97.044322,Nucl.Phys.A.935.28,Nucl.Phys.A.940.21,Int.J.Mod.Phys.E.22.1350081}, proton radioactivity\cite{Eur.Phys.J.A.55.58,Phys.Rev.C.96.034619}, $\alpha$ decay fine structure\cite{J.Phys.G:Nucl.Part.Phys.38.075101,Phys.Rev.C.86.024613}, heavy ion fusion and ternary fission\cite{Nucl.Phys.A.817.35,Nucl.Phys.A.922.191,J.Phys.G:Nucl.Part.Phys.41.105108,Phys.Rev.C.91.044603} and predict the $\alpha$ decay chains of superheavy nuclei\cite{Phys.Rev.C.84.024609,Phys.Rev.C.90.054614}. Considering the $2p$ radioactivity process being share the same theory as $\alpha$ decay and proton radioactivity i.e. barrier penetration processes, whether the CPPM can be extended to study $2p$ radioactivity or not is a desirable question. To this end, we extend the CPPM to systematically study the $2p$ radioactivity half-lives of proton-rich nuclei with 12\,\textless$Z$\textless\,36.

This article is organized as follows. In the next section, the theoretical framework of Coulomb and proximity potential model is presented. The results and discussion are shown in Sec. \ref{section 3}. Finally a summary is given in Sec. \ref{section 4} 

\section{Coulomb and proximity potential model}	
\label{zection 2}
The $2p$ radioactivity half-life is generally calculated by
\begin{equation}
T_{1/2} = \frac{ln2}{\lambda},
\end{equation}		
where $\lambda$ is the decay constant. It can be expressed as
\begin{equation}
{\lambda} = S_{2p} \nu \emph{P}.
\end{equation}
Here $ \nu $ is the assault frequency related to the harmonic oscillation frequency presented in the Nilsson potential\cite{Dan.Mat.-Fys.Medd.29.16}. It can be expressed as  
\begin{equation}
h\nu=\hbar\omega\simeq\frac{41}{A^{1/3}},
\end{equation} 
where \emph{h}, $\hbar$,  $\omega$, and $A$ are the Planck constant, reduced Plank constant, angular frequency, and mass number of parent nucleus, respectively. $S_{2p}=G^2[A/(A-2)]^{2n}\chi^2$ represents the preformation probability of the emitted two protons in the parent nucleus and which is obtained by the cluster overlap approximation with $G^2 = (2n)!/[2^{2n}(n!)^2]$\cite{Phys.Rev.C.43.R1513,Phys.Rep.12.201}. Here $n\,\approx(3Z)^{1/3}$-1 is the average principal proton oscillator quantum number\cite{Nucl.Stru.Vol.1} and $\chi^2$ is set as 0.0143 according to Ref.\cite{Phys.Rev.C.101.014301}. 

$P$ is the penetration probability, which can be calculated  by the semi-classical Wentzel-Kramers-Brillouin (WKB) approximation and expressed as 
\begin{equation}
\label{eq4}
P={\rm exp}\left[-2\int_{r_{in}}^{r_{out}} K(r)\,dr\right],
\end{equation}
where $K(r) = \sqrt{\frac{2 \mu}{\hbar^2}|{V(r)-Q_{2p}}|}$ is the wave number of the emitted two protons. $\mu = \frac{m_{2p}m_d}{m_{2p}+m_d}$ $\approx$ 938.3$\times$ 2 $\times$ $A_d/A$ MeV/c$^2$ denotes the reduced mass with $m_{2p}$ and $m_d$ being the masses of emitted two protons and daughter nucleus, respectively\cite{Chin.Phys.C.45.044110}. $r$ is the mass center distance between the emitted two protons and daughter nucleus.  
$r_{in}$ and $r_{out}$ are classical inner and outer turning points of potential barrier which satisfied the conditions $V(r_{in}) = V(r_{out}) = Q_{2p}$. $V(r)$ is the whole interaction potential between the emitted two protons and daughter nucleus including the nuclear potential $V_{N}(r)$, Coulomb potential $V_{C}(r)$ and centrifugal potential $V_{l}(r)$. It can be expressed as
\begin{equation}
V(r)=V_{N}(r)+V_{C}(r)+V_{l}(r).
\end{equation}
In the CPPM\cite{Pramana.58.611}, the nuclear potential was replaced by the proximity potential, which was firstly put forward by Blocki \textit{et al.} as a simple formalism in 1977\cite{Ann.Phys.105.427}. In this work, for instance, we choose the proximity potential formalism 1981 (Prox.81)\cite{Ann.Phys.132.53} to obtain the nuclear potential between the emitted two protons and daughter nuclei. In this set of proximity potential, the nuclear potential $V_N(r)$ can be expressed as 
\begin{equation}
V_N(r)=4 \pi \gamma b \bar {R}\Phi(\xi),
\end{equation}
where $\gamma$ = $\gamma_0[1-k_s1.7826\left(\frac{N-Z}{A}\right)^2]$ is the surface energy coefficient, with surface energy constant $\gamma_0=0.9517$ MeV/fm$^2$ and the surface asymmetry constant $k_s$ = 1.7826. Here \emph{N}, \emph{Z} and \emph{A} are the neutron number, proton number and mass number of parent nucleus, respectively. $b$ is the diffuseness of nuclear surface taken as unity, $\bar{R}$ is the mean curvature radius which can be written as 
\begin{equation}
\bar{R}=\frac{C_1C_2}{C_1+C_2}.
\end{equation}
Here $C_1$ and $C_2$ denote the matter radii of daughter nucleus and emitted two protons, receptively. They have the following form 
\begin{equation}
C_i=R_i\left[1-\left(\frac{b}{R_i}\right)^2\right]\quad(i=1,2),
\end{equation}
where $R_1$ and $R_2$ are the radii of daughter nucleus and emitted two protons, respectively. The nuclei radii\cite{J.Phys.G:Nucl.Part.Phys.26.1149} can be parameterized as $R_i = 1.28A_i^{1/3}-0.76+-0.8A_i^{-1/3} (i=1,2)$. For the universal function $\Phi(\xi)$, it is expressed as

\begin{small}
\begin{equation}
\label{}
{\Phi(\xi)=
				\left\{\begin{array}{ll}
				-1.7817+0.9270\xi+0.143\xi^2-0.09\xi^3             &\xi<0,\\
				-1.7817+0.9270\xi+0.01696\xi^2-0.05148\xi^3        &0\le\xi\le1.9475,\\
				-4.41e^{-\xi/0.7176}                               &\xi>1.9475\\ 
						\end{array}\right.}
\end{equation}
\end{small}
where $\xi = \frac{r-C_1-C_2}{b}$ represents the distance between the near surface of the daughter and emitted two protons.

The Coulomb potential $V_{C}(r)$ is hypothesized as the potential of a uniformly charged sphere with sharp radius $R$. It is expressed as
\begin{equation}
\label{}
V_{C}(r)=
\left\{\begin{array}{ll}
\frac{Z_1Z_2e^2}{2R}[3-(\frac{r}{R})], &r<R,\\
\frac{Z_1Z_2e^2}{r},         &r>R,\\
\end{array}\right.
\end{equation}
where $R=R_1+R_2$ is the separation radius. $Z_1$ and $Z_2$ are the proton number of daughter nucleus and emitted two protons, respectively.

For the centrifugal potential $V_l(r)$, because $l(l+1)\to(l+\frac{1}{2})^2$ is a necessary correction for one-dimensional problems\cite{J.Math.Phys.36.5431}, we choose the Langer modified form. It can be written as
\begin{equation}
V_l(r)=\frac{\hbar^2(l+\frac{1}{2})^2}{2\mu r^2},
\end{equation}
where \emph{l} is the obrital angular momentum taken away by the emitted two protons. The minimum orbital angular momentum $l_{min}$ can be obtained by the parity and angular momentum conservation laws.

\section{Results and discussion}
\label{section 3}
In order to describe the interaction potential between any two nuclei in the separation degree of freedom, based on the proximity force theorem, Blocki \textit{et al}. proposed the proximity potential for the first time in 1977\cite{Ann.Phys.105.427}. Hence, various nuclear proximity potentials have been widely applied to study the nuclear physics \cite{Eur.Phys.J.A.50.187,Nucl.Phys.A.915.70,Phys.Rev.C.90.064603,Phys.Rev.C.85.054612,Phys.Rev.C.81.064609}. In 2002, using the proximity potential to replace the nuclear potential, Santhosh \textit{et al}. proposed the Coulomb and proximity potential model (CPPM) to study the cluster radioactivity. Later on, the CPPM was more broadly used to investigate the $\alpha$ decay and proton radioactivity\cite{Eur.Phys.J.A.55.58,Phys.Rev.C.96.034619,Phys.Rev.C.84.024609,Phys.Rev.C.90.054614}. For the $2p$ radioactivity, it may share the same theory i.e. barrier penetration process as $\alpha$ decay, proton radioactivity and cluster radioactivity. In this work, we extend the CPPM to systematically study the $2p$ radioactivity half-lives of the nuclei with 12\,\textless\emph{Z}\textless\,36.

At first, we performed calculations on the $2p$ radioactivity half-lives of the true $2p$ radioactive nuclei of $^{19}$Mg, $^{45}$Fe, $^{48}$Ni, $^{54}$Zn and $^{67}$Kr, amounting 10 experimental datasets within the CPPM, all the detailed calculated results are presented in Table \ref{Table 1}. For comparison, the experimental $2p$ radioactivity half-lives and the calculated ones obtained by effective liquid drop model (ELDM)\cite{Phys.Lett.B.774.14}, generalized liquid drop model (GLDM)\cite{Phys.Rev.C.101.014301}, and Gamow-like model\cite{Chin.Phys.C.45.044110} are also listed in Table \ref{Table 1}. In this table, the first four columns represent the radioactive parent nucleus, the $2p$ radioactivity released energy $Q_{2p}$, the angular momentum $l$ taken away by the emitted two protons and the logarithmic form of experimental $2p$ radioactivity half-life denoted as log$T_{1/2}^{\rm{expt}}$, respectively. The last six columns represent the logarithmic form of calculated $2p$ radioactivity half-life denoted as log$_{10}T^{\rm{calc}}_{1/2}$ calculated by different theoretical models and empirical formulas, including CPPM, GLDM\cite{Phys.Rev.C.101.014301}, ELDM\cite{Phys.Lett.B.774.14}, Gamow-like model\cite{Chin.Phys.C.45.044110}, Sreeja formula\cite{Eur.Phys.J.A.55.33} and Liu formula\cite{Chin.Phys.C.45.024108}, respectively. From this table, we can clearly see that our calculated results using CPPM have the same magnitude with the ones obtained by using the above mentioned theoretical models and empirical formulas. In order to further demonstrate the degree of agreement, we plot the differences (log$_{10}T^{\rm{calc}}_{1/2}$ - log$_{10}T^{\rm{expt}}_{1/2}$) between the experimental values and the theoretical ones calculated by different theoretical models and empirical formulas in Fig. \ref{Fig 1}. From this figure, it is obviously seen that the deviations for $2p$ radioactive nuclei are almost among -1 $\to$ +1 except for the $^{54}$Zn ($Q_{2p}$ = 1.28MeV) and $^{67}$Kr. For the case of $^{54}$Zn ($Q_{2p}$ = 1.28MeV), it is not difficult to find that the calculated results within all of theoretical approaches mentioned above have a evident deviation with experimental value, meanwhile the calculated results are very close to the experimental value for $Q_{2p}$ = 1.48MeV. Due to the significant deviations between the experimental half-lives and the ones obtained by theoretical approaches mentioned above, we suspect the experimental data maybe not accurate enough, either the released energies or the half-lives. In addition to the defect of detecting the experimental data, it may be caused by deformation of daughter and parent nucleus which plays a momentous role in the $2p$ radioactivity half-live pointed out by Goigoux \textit{et al.}\cite{Phys.Rev.Lett.117.162501}. In addition, the three-body asymptotic behavior and configuration mixing are noticed as key factors for the $2p$ radioactivity\cite{Phys.Rev.Lett.117.162501,Nucl.Phys.A.767.13,Phys.Part.Nucl.40.674}. Therefore, it is crucial to improve CPPM with taking the factors mentioned above into account to make more reliable predictions in the future work.
\end{multicols}

\begin{center}
\tabcaption{The comparison between the calculated $2p$ radioactivity half-lives by using CPPM, GLDM, ELDM and Gamow-like model, two empirical formulas and the experimental ones. The experimental $2p$ radioactivity half-lives and released energy $Q_{2p}$ are taken from the corresponding references.}
\label {Table 1}
\footnotesize
\begin{tabular}{cccccccccc}
\hline \hline
\multicolumn{2}{c}{\multirow{3}{*}{}}&\multicolumn{7}{c}{log$\rm{_{10}^{cal}}\emph{T}_{1/2}$ (s)} \\			
\cline{4-10} 
{Nuclei}&$Q_{2p}$ (MeV) &$l$& EXPT&CPPM&GLDM\cite{Phys.Rev.C.101.014301}&ELDM\cite{Phys.Lett.B.774.14}&Gamow-like\cite{Chin.Phys.C.45.044110}&Sreeja\cite{Eur.Phys.J.A.55.33}&Liu\cite{Chin.Phys.C.45.024108}\\ \hline
 \noalign{\global\arrayrulewidth1.2pt}\noalign{\global\arrayrulewidth0.4pt} \multicolumn{10}{c}{\textbf{}}\\

			$^{19}$Mg& 0.75\cite{Phys.Rev.Lett.99.182501} & 0 & -11.40 \cite{Phys.Rev.Lett.99.182501}  & -12.17 & -11.79 & -11.72 & -11.46 & -10.66 & -12.03\\
			$^{45}$Fe& 1.10\cite{Eur.Phys.J.A.14.279}     & 0 & -2.40  \cite{Eur.Phys.J.A.14.279}      & -2.07  & -2.23  & -      & -2.09  & -1.25  & -2.21 \\
			$$       & 1.14\cite{Phys.Rev.Lett.89.102501} & 0 & -2.07  \cite{Phys.Rev.Lett.89.102501}  & -2.55  & -2.71  & -      & -2.58  & -1.66  & -2.64 \\ 
			$$       & 1.15\cite{Phys.Rev.C.72.054315}    & 0 & -2.55  \cite{Phys.Rev.C.72.054315}     & -2.71  & -2.87  & -2.43  & -2.74  & -1.80  & -2.79 \\ 
			$$       & 1.21\cite{Eur.Phys.J.A.48.179}     & 0 & -2.42  \cite{Eur.Phys.J.A.48.179}      & -3.33  & -3.50  & -      & -3.37  & -2.34  & -3.35 \\ 
			$^{48}$Ni& 1.29\cite{Phys.Rev.C.90.014311}    & 0 & -2.52  \cite{Phys.Rev.C.90.014311}     & -2.41  & -2.62  & -      & -2.59  & -1.61  & -2.59 \\ 
			$$       & 1.35\cite{Phys.Rev.C.72.054315}    & 0 & -2.08  \cite{Phys.Rev.C.72.054315}     & -3.03  & -3.24  & -      & -3.21  & -2.13  & -3.13 \\ 
			$^{54}$Zn& 1.28\cite{Phys.Rev.Lett.107.102502}& 0 & -2.76  \cite{Phys.Rev.Lett.107.102502} & -0.71  & -0.87  & -      & -0.93  & -0.10  & -1.01 \\ 
			$$       & 1.48\cite{Phys.Rev.Lett.94.232501} & 0 & -2.43  \cite{Phys.Rev.Lett.94.232501}  & -2.79  & -2.95  & -2.52  & -3.01  & -1.83  & -2.81 \\ 
			$^{67}$Kr& 1.69\cite{Phys.Rev.Lett.117.162501}& 0 & -1.70  \cite{Phys.Rev.Lett.117.162501} & -0.22  & -1.25  & -0.06  & -0.76  &  0.31  & -0.58 \\ 
\hline \hline
\end{tabular}
\end{center} 

\begin{center}
\includegraphics[width=10cm]{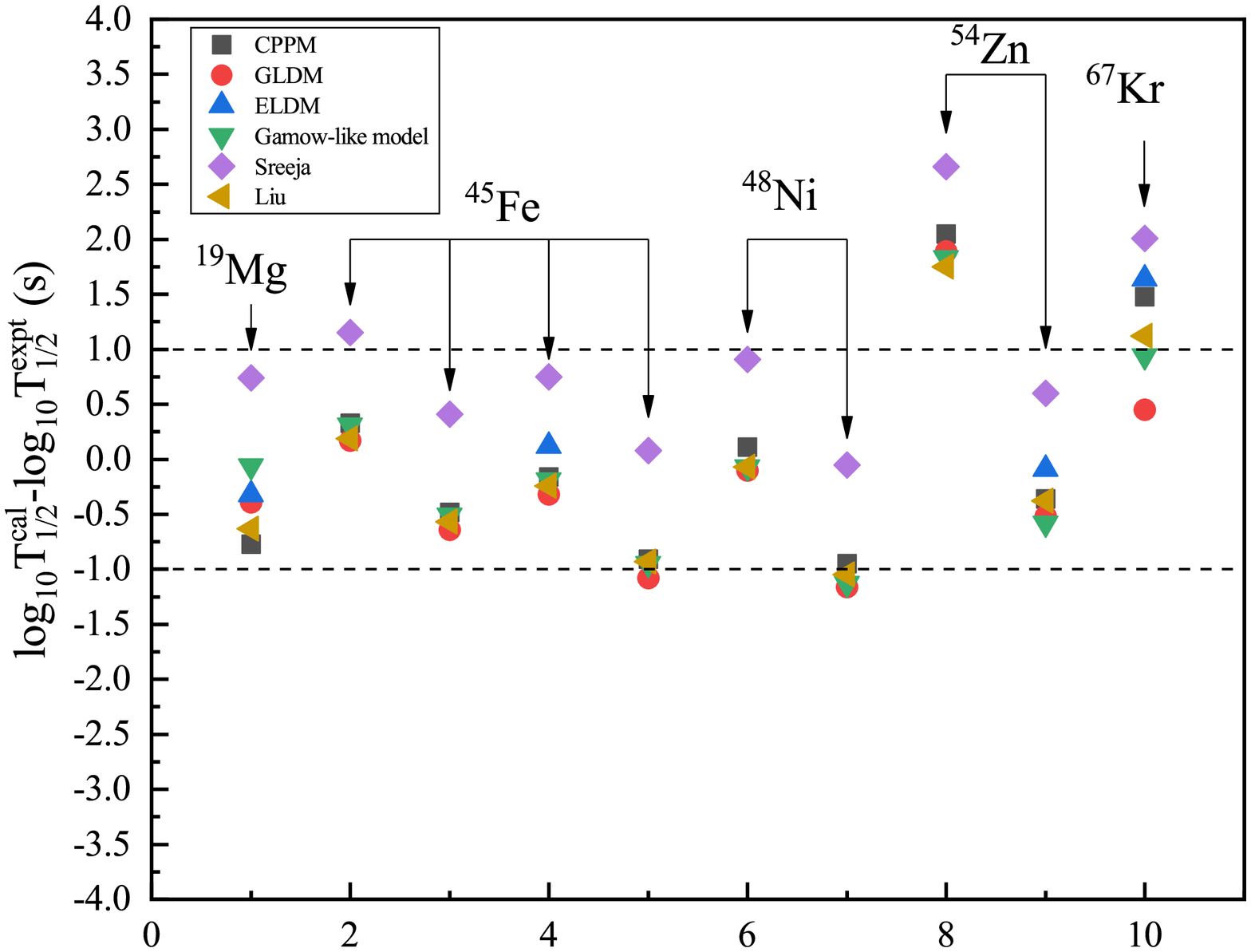}
\figcaption{\label{Figure1}(color online) Deviations between the experimental data and the calculated $2p$ radioactivity half-lives obtained by GLDM, ELDM, Gamow-like model, Sreeja formula and Liu formula.}
	\label {Fig 1}
\end{center}

\begin{multicols}{2}

Given the good agreement between the calculated results with CPPM and the experimental data, we use this model to predict the half-lives of possible $2p$ radioactive candidates with $Q_{2p}>$\,0 and $Q_p<$\,0, extracted from the 2016 Atomic Mass Evaluation (AME) mass table\cite{Chin.Phys.C.41.030002,Chin.Phys.C.41.030003,Chin.Phys.C.41.030001}. The predicted results are listed in Table \ref{Table 2}. In this table, the first three columns represent the radioactive parent nuclei, $Q_p$ and $Q_{2p}$, respectively. The predicted results that obtained by the CPPM, GLDM, ELDM, Gamow-like model, Sreeja formula and Liu formula are listed in the 4th to 9th column, respectively. From this table, we can clearly see the predicted results are quite different from the ones that obtained by using GLDM, ELDM, Gamow-like model, Sreeja formula and Liu formula while the values of the $Q_{2p}$ below 1 MeV. It maybe caused by penetration probability $P$. From Eq. \ref{eq4}, we can clearly obtain that the $P$ have a strong sensibility with $r_{in}$ and $r_{out}$ which are obtained by the conditions $V(r_{in}) = V(r_{out}) = Q_{2p}$. To illustrate the consistency of the predicted results using different models with CPPM, we plot the logarithmic predicted half-lives of possible candidates in Figure. \ref{Fig 2}. From this figure, it is obviously seen that the predicted results are in good agreement with the ones obtained by GLDM, ELDM and Gamow-like model except for $^{55}$Zn and $^{64}$Se, it may need future experiment to check this phenomenon.
\end{multicols}

\begin{center}
	\tabcaption{The comparison between the calculated $2p$ radioactivity half-lives by using CPPM, GLDM, ELDM and Gamow-like model, two empirical formulas and the experimental ones. The experimental $2p$ radioactivity half-lives and released energy $Q_{2p}$ are taken from the corresponding references.}
	\label {Table 2}
	\footnotesize
	\begin{tabular}{ccccccccc}
		\hline \hline
		\multicolumn{2}{c}{\multirow{3}{*}{}}&\multicolumn{6}{c}{log$\rm{_{10}^{pre}}\emph{T}_{1/2}$ (s)} \\			
		\cline{4-9} 
		{Nuclei}&$Q_p$ (MeV)&$Q_{2p}$ (MeV) & CPPM & GLDM\cite{Phys.Rev.C.101.014301} & ELDM\cite{Phys.Lett.B.774.14} & Gamow-like model\cite{Chin.Phys.C.45.044110} & Sreeja\cite{Eur.Phys.J.A.55.33} & Liu\cite{Chin.Phys.C.45.024108} \\ \hline
		\noalign{\global\arrayrulewidth1.2pt}\noalign{\global\arrayrulewidth0.4pt} \multicolumn{9}{c}{\textbf{}}\\
		
		$^{22}_{14}$Si & -0.94 & 1.28 & -13.73 & -13.30 & -13.32 & -13.25  & -12.30 & -13.74 \\
		$^{34}_{20}$Ca & -0.48 & 1.47 & -10.33 & -10.71 & -9.91  & -10.10  & -8.65  & -9.93  \\
		$^{39}_{22}$Ti & -0.84 & 0.76 & -1.24  & -1.34  & -0.81  & -0.91   & -0.28  & -1.19  \\
		$^{42}_{24}$Cr & -0.88 & 1.00 & -2.74  & -2.88  & -2.43  & -2.65   & -1.78  & -2.76  \\
		$^{49}_{24}$Ni & -0.59 & 0.49 & 10.23  & 14.46  & 14.64  & 14.54   & 12.78  & 12.43  \\
		$^{55}_{30}$Zn & -0.45 & 0.48 & 11.87  & 17.94  & -      & -       & -      & -      \\
		$^{60}_{32}$Ge & -0.62 & 3.63 & 11.47  & 13.55  & 14.62  & 14.24   & 12.40  & 12.04  \\
		$^{64}_{34}$Se & -0.49 & 0.46 & 15.04  & 24.44  & -      & -       & -      & -      \\
		\hline \hline	
	\end{tabular}
\end{center}
	
\begin{center}
	\includegraphics[width=10cm]{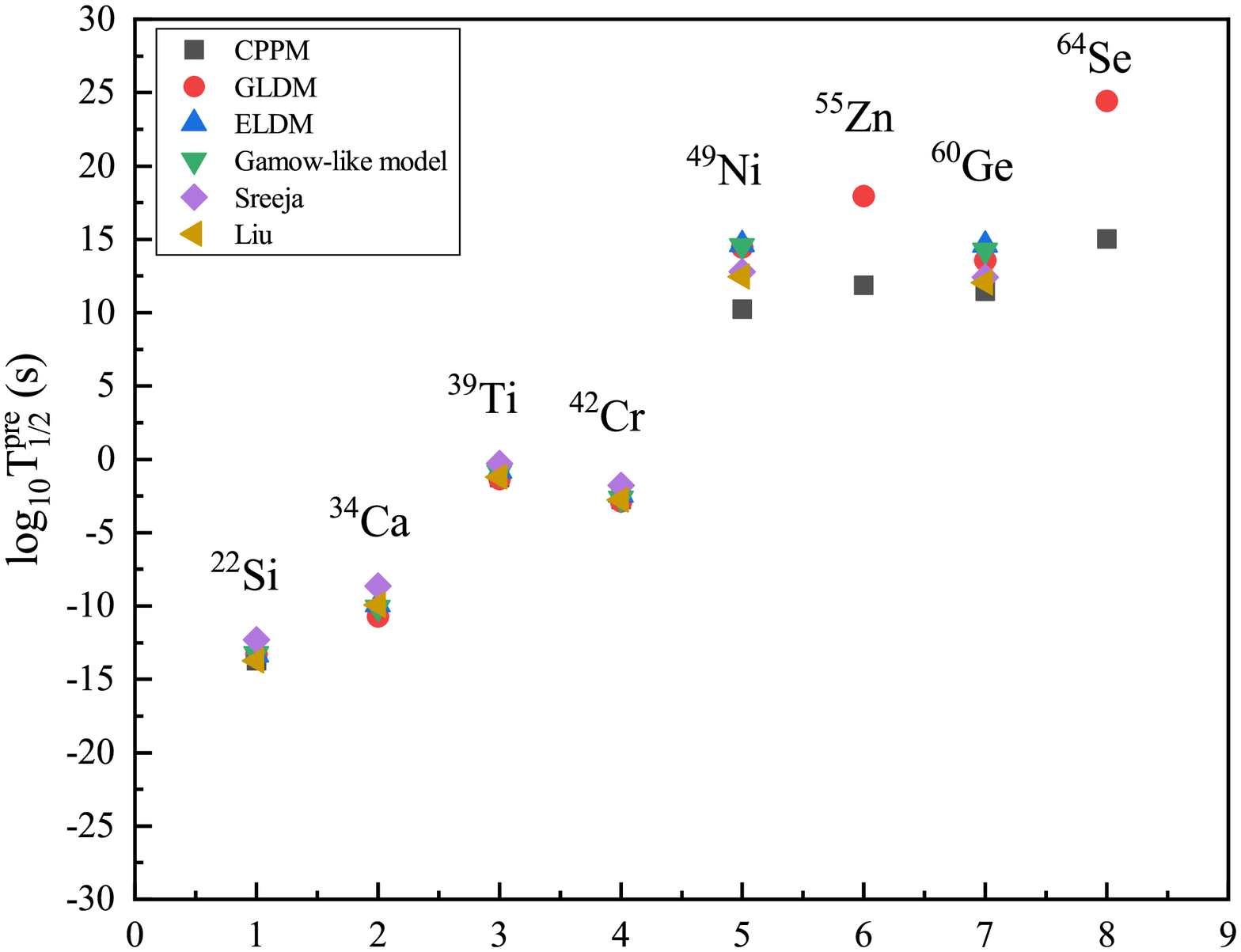}
	\figcaption{\label{Figure2} (color online) The predicted half-lifes witnin CPPM comparing with the ones by using the ELDM, GLDM, Gamow-like model, Sreeja formula and Liu formula.}
	\label {Fig 2}
\end{center}

\begin{multicols}{2}
\section{Summary}
\label{section 4}
In this work, we extend the Coulomb and proximity potential (CPPM) to systematically investigate the half-lives of two-proton ($2p$) radioactive nuclei including $^{19}$Mg, $^{45}$Fe, $^{48}$Ni, $^{54}$Zn and $^{67}$Kr. In calculations, the proximity potential is chosen as proximity potential formalism 1981 (Prox.81), meanwhile, the preformation probability ($S_{2p}$) of the two protons in the parent nucleus is taken into consideration. The calculated results within the CPPM are in good agreement with the experimental data, other theoretical models and empirical formulas. In addition, we predict the half-lives of possible $2p$ radioactive candidates. It may be provided a theoretical reference for the future experiments.
\end{multicols}
\vspace{-1mm}
\centerline{\rule{80mm}{0.1pt}}
\vspace{2mm}

\begin{multicols}{2}

\end{multicols}

\clearpage
\end{CJK*}

\begin{thebibliography}{90}

\bibitem{Sov.Phys.JETP.11.812} Y. B. Zel'dovich, Sov. Phys. JETP \textbf{11} 812 (1960)

\bibitem{Nucl.Phys.19.482} V. I. Goldansky, Nucl. Phys \textbf{19} 482 (1960)

\bibitem{Nucl.Phys.27.684} V. I. Goldansky, Nucl. Phys \textbf{27} 684 (1961)

\bibitem{Nucl.Phys.23.366} V. I. Goldansky, Nucl. Phys \textbf{23} 366 (1976)

\bibitem{Phys.Rev.C.90.054326} Q. Zhao, J. M. Dong, J. L. Song and W. H. Long, Phys. Rev. C \textbf{90} 054326 (2014)

\bibitem{Phys.Rev.Lett.111.139903} E. Olsen,  M. Pf\"utzner, N. Birge \textit{et al} Phys. Rev. Lett \textbf{111} 139903 (2013)

\bibitem{Eur.Phys.J.A.54.65} O. A. P. Tavares and E. L. Medeiros, Eur. Phys. J. A \textbf{54} 65 (2018)

\bibitem{Phys.Rev.C.68.054005} L. V. Grigorenko and M. V. Zhukov, Phys. Rev. C \textbf{68} 054005 (2003)

\bibitem{Phys.Rev.C.79.054330} J. M. Dong, H. F. Zhang, and G. Royer, Phys. Rev. C \textbf{79} 054330 (2009)

\bibitem{Int.J.Mod.Phys.E.30.2150074} H. M. Liu, Y. T. Zou, X. Pan \textit{et al} Int. J. Mod. Phys. E \textbf{30} 2150074 (2021)

\bibitem{Commun.Theor.Phys.73.075302} X. Pan, Y. T. Zou \textit{et al} Commun. Theor. Phys \textbf{73} 075302 (2021)

\bibitem{Commun.Theor.Phys.73.075301} Y. Z. Wang, J. P. Cui \textit{et al} Commun. Theor. Phys \textbf{73} 075301 (2021)

\bibitem{Chin.Phys.C.45.104102} Y. T. Zou, X. Pan , X. H. Li \textit{et al} Chin. Phys. C \textbf{45} 104102 (2021)

\bibitem{Chin.Phys.C.45.044110} H. M. Liu, X. Pan, Y. T. Zou \textit{et al} Chin. Phys. C \textbf{45} 044110 (2021)

\bibitem{Eur.Phys.J.A.55.33} I. Sreeja and M. Balasubramaniam, Eur. Phys. J. A \textbf{55} 33 (2019)

\bibitem{Phys.Rev.C.96.034619} K. P. Santhosh and I. Sukumaran \textit{et al} Phys. Rev.C \textbf{96} 034619 (2017)

\bibitem{Phys.Rev.C.100.054332} B. A. Brown, B. Blank and J. Giovinazzo  \textit{et al}  Phys. Rev. C \textbf{100} 054332 (2019)

\bibitem{Eur.Phys.J.A.55.214} J. L. Chen, J. Y. Xu, J. G. Deng  \textit{et al}  Eur. Phys. J. A \textbf{55} 214 (2019)

\bibitem{Phys.Rev.C.64.054002} L. V. Grihorenko, R. C. Johnson \textit{et al}  Phys. Rev. C \textbf{64} 054002 (2001)

\bibitem{Rep.Prog.Phys.71.046301} B. Blank and M. Ploszajczak, Rep. Prog. Phys \textbf{71} 046301 (2008)

\bibitem{Prog.Part.Nucl.Phys.59.418} L. S. Ferreira, M. C. Lopea and E. Maglione, Prog. Part. Nucl. Phys \textbf{59} 418 (2007)

\bibitem{Phys.Rev.C.69.054311} A. Kruppa and W. Nazarewicz, Phys. Rev. C \textbf{69} 054311 (2004)

\bibitem{Phys.Rev.150.836} W. Whaling, Phys. Rev \textbf{150} 836 (1966)

\bibitem{Phys.Rev.Lett.74.860} R. A. Kryger, A. Azhari, M. Hellstr\"om \textit{et al} Phys. Rev. Lett \textbf{74} 860 (1995)

\bibitem{Phys.Rev.C.17.1929} G. J. KeKelis, M. S. Zisman, D. K. Scott \textit{et al} Phys. Rev. C \textbf{17} 1929 (1978)

\bibitem{Rev.Mod.Phys.84.567}  M. Pf\"utzner, M. Karny, L. V. Grihorenko \textit{et al} Rev. Mod. Phys \textbf{84} 567 (2012)

\bibitem{Eur.Phys.J.A.14.279}M. Pf\"utzner, E. Badura \textit{et al} Eur. Phys. J. A \textbf{14} 279 (2002)

\bibitem{Phys.Rev.Lett.89.102501} J. Giovinazzo, B. Blank \textit{et al} Phys. Rev. Lett \textbf{89} 102501 (2002)	

\bibitem{Phys.Rev.Lett.94.232501} B. Blank, A. Bey, G. Canchel \textit{et al} Phys. Rev. Lett \textbf{94} 232501 (2005)

\bibitem{Phys.Rev.Lett.107.102502} P. Ascher, L. Audirac, N. Adimi \textit{et al} Phys. Rev. Lett \textbf{107} 102502 (2011)

\bibitem{Phys.Rev.Lett.99.182501} I. Mukha, K. S\"ummerer, L. Acosta \textit{et al} Phys. Rev. Lett \textbf{99} 182501 (2007)	

\bibitem{Phys.Rev.Lett.84.1116} B. Blank, A. Bey \textit{et al} Phys. Rev. Lett \textbf{84} 1116 (2000)

\bibitem{Phys.Rev.C.83.061303} M. Pomorski, M. Pf\"utzner, W. Dominik \textit{et al} Phys. Rev. C \textbf{83} 061303(R) (2011)

\bibitem{Phys.Rev.Lett.117.162501} T. Goigoux, P. Ascher, B. Blank \textit{et al} Phys. Rev. Lett \textbf{117} 162501 (2016)

\bibitem{J.Phys.G:Nucl.Part.Phys.31.S1509} J. Giovinazzo, J. Phys. G: Nucl. Part. Phys \textbf{31} S1509 (2005)

\bibitem{Eur.Phys.J.A.48.179} L. Audirac, P. Ascher, B. Blank1 \textit{et al} Eur. Phys. J. A \textbf{48} 179 (2012)

\bibitem{Phys.Rev.C.53.214} W. E. Ormand, Phys. Rev. C \textbf{53} 214 (1996)

\bibitem{Phys.Rev.C.54.1240} B. J. Cole, Phys. Rev. C \textbf{54} 1240 (1996)

\bibitem{Phys.Rev.C.56.1866} B. J. Cole, Phys. Rev. C \textbf{56} 1866 (1997)

\bibitem{Phys.Rev.C.58.2813} B. J. Cole, Phys. Rev. C \textbf{58} 2813 (1998)

\bibitem{Phys.Rev.C.59.726} B. J. Cole, Phys. Rev. C \textbf{59} 726 (1999)

\bibitem{Phys.Lett.B.774.14} M. Gonalves, N. Teruya,  O. A. P. Tavares \textit{et al} Phys. Lett. B \textbf{774} 14 (2017)

\bibitem{Phys.Rev.C.101.014301} J. P. Cui, Y. H. Gao \textit{et al} Phys. Rev. C \textbf{101} 014301 (2020)\\
								J. P. Cui, Y. H. Gao \textit{et al} Phys. Rev. C \textbf{104} 029902(E) (2021) 

\bibitem{Nucl.Phys.56.86} V. M. Galitsky and V. F. Cheltsov, Nucl. Phys \textbf{56} 86 (1976)

\bibitem{Phys.Rev.Lett.110.222501} E. Olsen \textit{et al} Phys. Rev. Lett \textbf{110} 222501 (2013)

\bibitem{Phys.At.Nucl.56.460} B. V. Danilin and M. V. Zhukov, Phys. At. Nucl \textbf{56} 460 (1993) 

\bibitem{Phys.Rev.C.63.034607} V. Vasilevsky \textit{et al} Phys. Rev. C \textbf{63} 034607 (2001)

\bibitem{Nucl.Phys.A.765.370} P. Descouvemount \textit{et al} Nucl. Phys. A \textbf{765} 370 (2006)

\bibitem{Phys.Rev.C.76.014008} L. V. Grigorenko and M. V. Zhukov, Phys. Rev. C \textbf{76} 014008 (2007)

\bibitem{Phys.Rev.C.78.034004} E. Garrido \textit{et al} Phys. Rev. C \textbf{78} 034004 (2008)

\bibitem{Phys.Rev.C.77.064305} R. \'Alvarez-Rodr\'iguez, A. S. Jensen \textit{et al} Phys. Rev. C \textbf{77} 064305 (2008)

\bibitem{Phys.Rev.C.82.034001}R. \'Alvarez-Rodr\'iguez, A. S. Jensen \textit{et al}  Phys. Rev. C \textbf{82} 034001 (2010)

\bibitem{Acta.Phys.Pol.B.42.545} B. Blank \textit{et al} Phys. Rev. C \textbf{42} 545 (2011)

\bibitem{Phys.Rev.Lett.85.22} L. V. Grigorenko, R. C. Johnson, I. G. Mukha \textit{et al} Phys. Rev. Lett \textbf{85} 22 (2000)

\bibitem{Chin.Phys.C.45.024108} H. M. Liu, Y. T. Zou, X. Pan \textit{et al} Chin. Phys .C \textbf{45} 024108 (2021)

\bibitem{Ann.Phys.105.427} J. Blocki, J. Randrup and W. J. \ifmmode \acute{S}\else \'{S}\fi{}wia\ifmmode \mbox{\c{}}\else \c{}\fi{}tecki \textit{et al}  Ann. Phys \textbf{105} 427 (1977)

\bibitem{Nucl.Phys.A.850.34} K. P. Santhosh, S. Sabina and J. G. Joseph, Nucl. Phys. A \textbf{850} 34 (2011)

\bibitem{Nucl.Phys.A.882.49} K. P. Santhosh, S. Sabina \textit{et al} Nucl. Phys. A \textbf{882} 49 (2012)

\bibitem{Nucl.Phys.A.889.29} K. P. Santhosh, B. Priyanka \textit{et al} Nucl. Phys. A \textbf{889} 29 (2012)

\bibitem{Nucl.Phys.A.929.20} K. P. Santhosh and B. Priyanka, Nucl. Phys. A \textbf{929} 20 (2014)

\bibitem{Nucl.Phys.A.935.28} K. P. Santhosh, I. Sukumaran and B. Priyanka, Nucl. Phys. A \textbf{935} 28 (2015)

\bibitem{Nucl.Phys.A.922.191} K. P. Santhosh and  V. B. Jose, Nucl. Phys. A \textbf{922} 191 (2014)

\bibitem{J.Phys.G:Nucl.Part.Phys.41.105108} K. P. Santhosh, S. Krishnan \textit{et al} J. Phys. G: Nucl. Part. Phys \textbf{41} 105108 (2014)

\bibitem{Can.J.Phys.95.31}K. P. Santhosh and I. Sukumaran, Can. J. Phys \textbf{95} 31 (2016)

\bibitem{Phys.Rev.C.62.044610} W. D. Myers and W. J. \ifmmode \acute{S}\else \'{S}\fi{}wia\ifmmode \mbox{\c{}}\else \c{}\fi{}tecki, Phys. Rev. C \textbf{62} 044610 (2000)

\bibitem{Nucl.Phys.A.438.450} Y. J. Shi and W. J. Swiatecki, Nucl. Phys. A \textbf{438} 450 (1985)

\bibitem{Pramana.76.6} I. Dutt, Pramana. J .Phys \textbf{76} 6 (2011)

\bibitem{Phys.Let.65B.1} P. R. Christensen and A. Winther, Phys. Let \textbf{65B} 1 (1976)

\bibitem{Phys.Rev.C.81.044615} I. Dutt and R. K. Puri, Phys. Rev. C \textbf{81} 044615 (2010)

\bibitem{Pramana.58.611} K. P. Santhosh and A. Joseph, Pramana \textbf{58} 611 (2002)

\bibitem{Phys.Rev.C.93.024612} O. N. Ghodsi and A. Daei-Ataollah, Phys. Rev. C \textbf{93} 024612 (2016)

\bibitem{Phys.Rev.C.97.044322} J. G. Deng, J. C. Zhao, P. C. Chu \textit{et al} Phys. Rev. C \textbf{97} 044322 (2018)

\bibitem{Eur.Phys.J.A.51.122} Y. J. Yao,  G. L. Zhang,  W. W. Qu \textit{et al} Eur. Phys. J. A \textbf{51} 122 (2015)

\bibitem{Int.J.Mod.Phys.E.22.1350081} K. P. Santhosh and B. Priyanka, Int. J. Mod. Phys. E \textbf{22} 1350081 (2013)

\bibitem{Nucl.Phys.A.940.21} K. P. Santhosh and B. Priyanka, Nucl. Phys. A \textbf{940} 21 (2015)

\bibitem{Eur.Phys.J.A.55.58} J. G. Deng, X. H. Li, J. L. Chen \textit{et al} Eur. Phys. J. A \textbf{55} 58 (2019)

\bibitem{Phys.Rev.C.86.024613} K. P. Santhosh and J. G. Joseph, Phys. Rev. C \textbf{86} 024613 (2012)

\bibitem{J.Phys.G:Nucl.Part.Phys.38.075101} K. P. Santhosh, J. G. Joseph \textit{et al} J. Phys. G: Nucl. Part. Phys \textbf{38} 075101 (2011)

\bibitem{Nucl.Phys.A.817.35} K. P. Santhosh, V. B. Jose \textit{et al} Nucl. Phys. A \textbf{817} 35 (2009)

\bibitem{Phys.Rev.C.91.044603} K. P. Santhosh, S. Krishnan \textit{et al} Phys. Rev. C \textbf{91} 044603 (2015)

\bibitem{Phys.Rev.C.84.024609} K. P. Santhosh, B. Priyanka \textit{et al} Phys. Rev. C \textbf{84} 024609 (2011)

\bibitem{Phys.Rev.C.90.054614}K. P. Santhosh and B. Priyanka, Phys. Rev. C \textbf{90} 054614 (2014)

\bibitem{Dan.Mat.-Fys.Medd.29.16} S. G. Nilsson, Dan. Mat .Fys. Medd \textbf{29} 16 (1955)
 
\bibitem{Phys.Rev.C.43.R1513} B. A. Brown, Phys. Rev. C \textbf{43} R1513 (1991)	

\bibitem{Phys.Rep.12.201} N. Anyas-Weiss \textit{et al} Phys. Rep \textbf{12} 201 (1991)

\bibitem{Nucl.Stru.Vol.1} A. Bohr,  B. R. Mottelson \textit{et al} Nucl. Stru (Benjiamin, New York) Vol \textbf{1} (1969)

\bibitem{Ann.Phys.132.53} J. Blocki and W. J. \ifmmode \acute{S}\else \'{S}\fi{}wia\ifmmode \mbox{\c{}}\else \c{}\fi{}tecki, Ann. Phys \textbf{132} 53 (1981)

\bibitem{J.Phys.G:Nucl.Part.Phys.26.1149} G. Rover, J. Phys. G: Nucl. Part. Phys \textbf{26} 1149 (2000)

\bibitem{J.Math.Phys.36.5431} J. J. Morehead, J. Math. Phys \textbf{36} 5431 (1955)

\bibitem{Eur.Phys.J.A.50.187} C. L. Guo and G. L. Zhang, Eur. Phys. J. A \textbf{50} 187 (2014)

\bibitem{Nucl.Phys.A.915.70} L. Zheng, G. L. Zhang \textit{et al} Nucl. Phys. A \textbf{915} 70 (2013)

\bibitem{Phys.Rev.C.90.064603} W. W. Qu, G. L. Zhang \textit{et al} Phys. Rev. C \textbf{90} 064603 (2014)

\bibitem{Phys.Rev.C.85.054612} R. Kumar and M. K. Sharma \textit{et al} Phys. Rev. C \textbf{85} 054612 (2014)

\bibitem{Phys.Rev.C.81.064609} I. Dutt, and R. K. Puri, Phys. Rev. C \textbf{81} 064609 (2010)

\bibitem{Nucl.Phys.A.767.13} J. Rotureau, J. Okolowicz and M. Ploszajczak, Nucl. Phys. A \textbf{767} 13 (2006)

\bibitem{Phys.Part.Nucl.40.674} L. V. Grigorenko, Phys. Part. Nucl \textbf{40} 674 (2009)

\bibitem{Phys.Rev.C.72.054315} C. Dossat \textit{et al} Phys. Rev. C \textbf{72} 054315 (2015)

\bibitem{Phys.Rev.C.90.014311} M. Pomorski \textit{et al} Phys. Rev. C \textbf{90} 014311 (2014)

\bibitem{Chin.Phys.C.41.030002} W. Huang, G. Audi, M. Wang \textit{et al} Chin. Phys. C \textbf{41} 030002 (2017)

\bibitem{Chin.Phys.C.41.030003} M. Wang , G. Audi, F. C Kondev \textit{et al} Chin. Phys. C \textbf{41} 030003 (2017)

\bibitem{Chin.Phys.C.41.030001} G. Audi, F. C. Kondev, M. Wang \textit{et al} Chin. Phys. C \textbf{41} 030001 (2017)






\vspace{3mm}
\end{thebibliography}
\end{document}